\documentclass[review]{elsarticle}
\usepackage{lineno,hyperref}
\usepackage{color}
\usepackage[dvipsnames]{xcolor}
\usepackage[normalem]{ulem}
\pdfoutput=1
\modulolinenumbers[5]
\journal{Journal of \LaTeX\ Templates}

\begin{document}
\begin{frontmatter} 
\title{$\eta$-pairing in a two-band model of spinless fermions}
\tnotetext[mytitlenote]{$\eta$-pairing  in a two-band model of spinless fermions}
\author{Igor N. Karnaukhov}
\address{G.V. Kurdyumov Institute for Metal Physics, 36 Vernadsky Boulevard, 03142 Kiev, Ukraine}
\fntext[myfootnote]{karnaui@yahoo.com}

\begin{abstract}
We study the two-band model of spinless fermions in which itinerant fermions  interact with localized fermions through the two-particle hybridization. In 1D version, the model has exact solution using the Bethe ansatz. 
It has been shown that accounting for two-particle hybridization reduces the repulsive interaction between itinerant fermions.
 In the case of strong interaction, the effective interaction between itinerant fermions is attractive, and $\eta$-pairing of spinless fermions is realized.
The proposed pairing mechanism via two-particle hybridization can lead to $p$-superconducting states with $\eta$-pairing.
$\eta$-pairing of spinless fermions could explain the phenomenon of high-temperature superconductivity experimentally observed in hydrogen-rich materials at high pressures.
 \end{abstract}
 \begin{keyword}
\texttt  pairing \sep hybridization \sep high-temperature superconductivity
\end{keyword}
\end{frontmatter}

\section{Introduction}

Forty years ago, Bednorz and M\"{u}ller \cite{1} discovered the phenomenon of high-temperature superconductivity. Progress has been made in producing a wide variety of compounds that exhibit superconducting properties at various, even quite high, temperatures. The maximal value of superconducting transition temperature $\sim 250K$ has been achieved in LaH$_{10}$ \cite{2,3,4} at high pressure. There is very little time left to realize superconductivity at room temperature. Despite obvious progress in obtaining new compounds with high-temperature superconductivity properties, we have not advanced one step in understanding the nature of high-temperature conductivity. None of the numerous theories proposed during this period \cite{a2,a3,a4} have stood the test of time or become effective in understanding the nature of high-temperature superconductivity.  Cooper's electron-phonon pairing mechanism \cite{BCS,a1} remains the only viable explanation for the nature of superconductivity. Unfortunately, it is unable to explain the high temperatures of the superconducting transition observed in high-temperature superconductors.

Direct interaction between fermions does not allow them to form pairs with subsequent condensation. 
Indirect interactions (as occurs with Cooper pairing via phonons) can lead to effective attraction between  electrons, and this interaction must be electron-electron (to explain the high superconducting transition temperatures). Such a mechanism must be fairly universal (like Cooper pairing, for example), since high-temperature superconductivity is realized in both complex and relatively simple two-component compounds. This paper focuses on the interaction between itinerant and localized spinless fermions in the form of  the two-particle hybridization between them. This type of interaction has not yet been considered or studied in terms of explaining the fermion pairing mechanism. While not claiming to provide a comprehensive explanation for the nature of high-temperature superconductivity, it is possible that this pairing mechanism could be realized in some compounds. Moreover, numerical calculations of the electronic states in hydrogen sulphide H$_3$S, palladium hydride PdH demonstrate a sharp increase in the hybridization of H-S and Pd-H electronic states near the Fermi energy \cite{2}.

\section{The model Hamiltonian}

The Hamiltonian of the model defines the two-band model in which itinerant $s$- and localized $d$-spinless fermions interact via intra-band density-density interactions, and the inter-band interaction between fermions is determined by the two-particle $s$-$d$ hybridization. The Hamiltonian  has the following form  ${\cal H}={\cal H}_s+{\cal H}_d+{\cal H}_{s-d}$:
\begin{eqnarray}
&&{\cal H}_s=- \sum_{{j}}^{L-1}(c^\dagger_{{j}}c_{{j+1}} +c^\dagger_{{j+1}}c_{{j}})
+2J\sum_{j=1}^{L-1}  n^s_{{j}}n^s_{{j+1}},\nonumber \\&&
{\cal H}_d=\varepsilon \sum_{j=1}^{L}  n^d_{{j}}+2I \sum_{j=1}^{L-1}  n^d_{{j}}n^d_{{j+1}},\nonumber \\&&    
{\cal H}_{s-d}=g \sum_{j=1}^{L-1} (c^\dagger_{{j}}c^\dagger_{{j+1}}d_{{j}}d_{{j+1}}+
d^\dagger_{{j+1}}d^\dagger_{{j}}c_{{j+1}}c_{{j}}),\nonumber \\
\label{eq:1}
\end{eqnarray}
where $c^\dagger_{{j}},c_{{j}}$ and  $d^\dagger_{{j}},d_{{j}}$  are the Fermi operators for $s$- and $d$-spinless fermions defined on the site ${j}$, $\varepsilon$ is an one-particle energy of $d$-fermions, which define the position of flat $d$-band relative to $s$-conduction band, $n^s_{j}=c^\dagger_{j}c_{j}$, $n^d_{j}=d^\dagger_{j}d_{j}$ are the density operators for $s$- and $d$-spinless fermions. $J$ and $I$ determine  the interactions of spinless fermions located at the nearest-neighboring sites within  bands. Parameter $g$  determines the magnitude of the two-particle $s$-$d$ hybridization of fermions located at the nearest neighboring sites.
The Hamiltonian conserves the total  number of fermions $\sum_j (n^d_{j}+n^s_{j})$. 
 $L$ is the total number of the lattice sites. 

Let us study the model described by the Hamiltonian (\ref{eq:1}). The one-particle energy of a localized fermion $\varepsilon$ is located outside the conduction band of $s$-fermions, below the band bottom (see Fig. 1). 
Hybridization between fermions occurs when the law of conservation of energy is satisfied, so when a level lies outside the band, the state corresponding to the energy of this level does not hybridize with the band states.
In this case, one-particle hybridization between $s$- and $d$-fermions is forbidden, and the two-particle on-site density-density interaction $\sim n^s_j n^d_j$ takes place. Since the one-particle state of the $d$-fermions is frozen, $n^d_j=1$ for such a configuration of fermion states. This interaction reduces to a one-particle one, shifting the energy of $s$-fermions. When the energy corresponding to one $d$-fermions in the state with two $d$-fermions located at the nearest neighboring sites $\varepsilon_2/2$ lies inside the conduction band of $s$-fermions (see in Fig.1), a two-particle hybridization between $s$- and $d$-fermions takes place. The only two-particle short-range interaction between $s$- and $d$-fermions in this case is the two-particle hybridization defined in the Hamiltonian  (\ref{eq:1}). 

The proposed model has two unique properties: first, it has an exact solution in one-dimensionality; second, in the case of strong interactions, it describes the $\eta$-pairing of itinerant fermions within the $p$-superconducting state. We will discuss these features of the model in detail below.

\begin{figure}[tp]
      \centering{\leavevmode}
\begin{minipage}[h]{.75\linewidth}
\center{
\includegraphics[width=\linewidth]{ 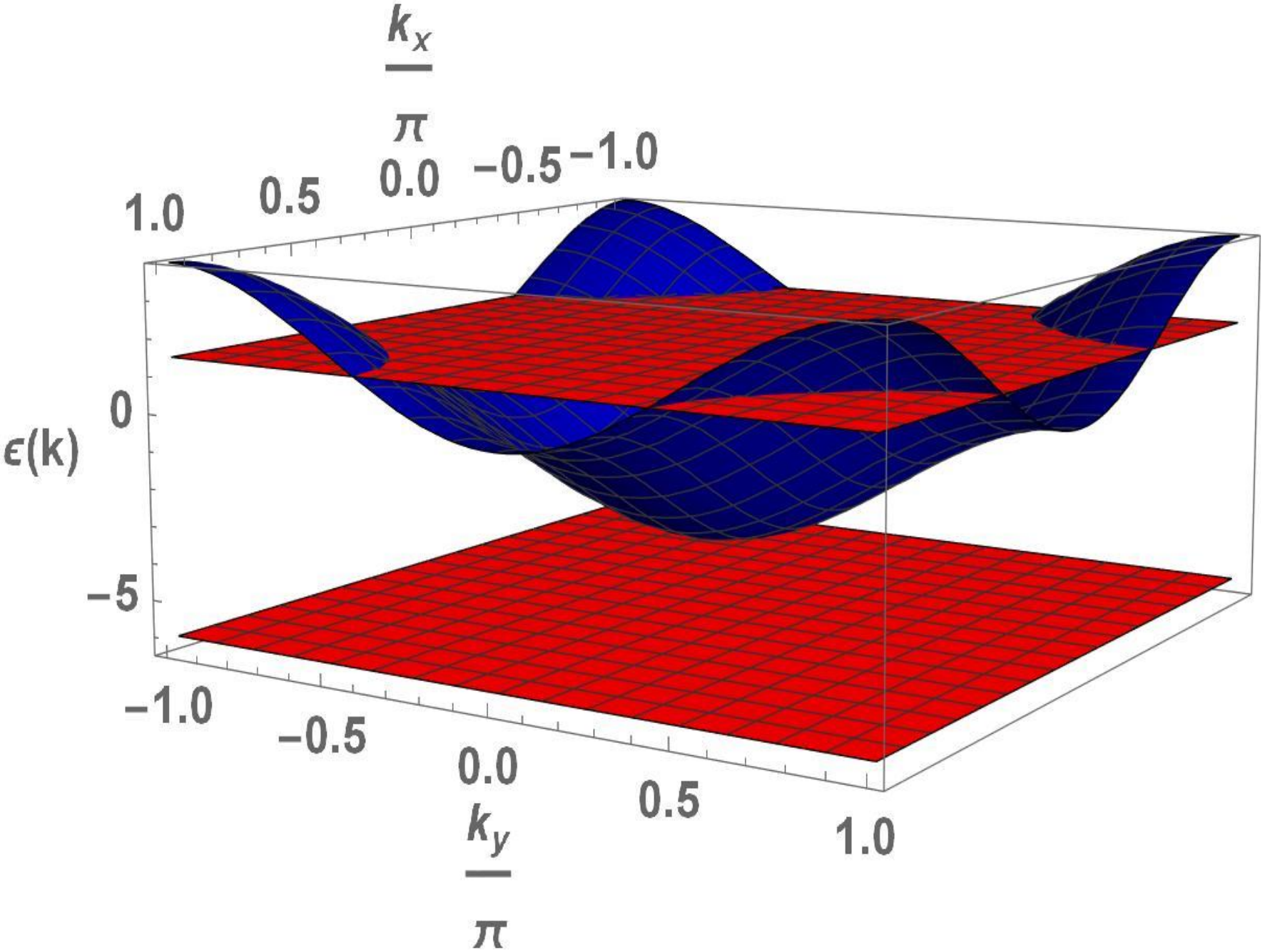} 
                  }
    \end{minipage}
\caption{The  one-particle energies of spinless fermions in the 2D model are shown: low-energy (read, with energy $\varepsilon=-6$) and high-energy (read, with energy $\varepsilon_2/2=\varepsilon+I=1\frac{1}{2}$) flat bands of the spinless $d$-fermions ($\varepsilon_2= 2\varepsilon+2I$ is the energy of two $d$-fermions located in the nearest neighboring sites), $\varepsilon (\textbf{k})=-2\cos k_x-2\cos k_y$ is the energy of $s$-fermion (blue, $\textbf{k}=\{k_x,k_y\}$ is the momentum of $s$-fermion).
} 
\label{fig:1}
\end{figure}

\section{Solution of the 1D model}

We start with one-particle problem, the wave function defines two states of spinless fermions:  $\psi(j) c^\dagger_{j}$, $\varphi(j) d^\dagger_{j}$. The amplitudes $\psi(j)$, $\varphi(j)$ satisfy the following equations:
\begin{eqnarray}
&&E \psi(j) =- \psi(j+{1})-\psi(j-{1}), \nonumber\\
&& (E-\varepsilon)\varphi(j)=0. 
\label{eq:2}
\end{eqnarray}

Eqs~(\ref{eq:2}) determine the energy of a conduction fermion, $E=\epsilon(k)=-2\cos k$, $k$ is the momentum of $s$-fermion, the energy of a localized $d$-fermion is equal to $E=\varepsilon$. One-particle state of $d$-fermions is frozen, it does not hybridize with conduction fermion and does not change its dispersion (see the energies of  the fermion states in Fig.1).

The eigenvector of the Hamiltonian (\ref{eq:1}) for two particles in the conduction band has the following form
\begin{eqnarray}
\vert \Psi>=\sum_{j_1,j_2} \psi(j_1,j_2)c^\dagger_{j_1}c^\dagger_{j_2}\vert0>
 +\sum_{j_1,j_2}  \varphi(j_1,j_2)d^\dagger_{j_1}d^\dagger_{j_2}\vert 0>, 
\label{eq:3}
\end{eqnarray}
where the amplitudes $\psi(j_1,j_2)$ and $\varphi(j_1,j_2)$ satisfy the Schr\"{o}dinger equation for $j_1\neq j_2 \pm 1$
\begin{eqnarray}
&&
E \psi (j_1,j_2)+
\psi(j_1+{1},j_2) + \psi(j_1-{1},j_2) +\psi(j_1,j_2+{1})+\psi(j_1,j_2-{1})=0,
\nonumber\\&&
 (E -2\varepsilon)\varphi (j_1,j_2)=0.
 \label{eq:4}
 \end{eqnarray} 
 
Two-particle state of the conduction fermions has energy $E= \epsilon(k_1)+\epsilon (k_2)$, here  $k_1$ and $k_2$ are the momenta of the $s$-fermions, located at the sites $j_1$ and $j_2$. Localized fermions located at the nearest neighboring sites  are hybridized with  conduction fermions. We write the equations for the amplitudes of the wave function $\psi(j_1,j_2)$ and $\varphi(j_1,j_2)$  at $j_1= j_2  \pm 1$
\begin{eqnarray}
&&
-[\psi(j,j) + \psi(j \pm 1,j \pm 1)] =2J\psi (j,j \pm 1)+g\varphi(j,j \pm 1),\nonumber \\&&
   [\epsilon (k_1)+ \epsilon (k_2)-\varepsilon_2]\varphi (j,j \pm 1)=
g\psi (j,j,\pm 1),
 \label{eq:5}
 \end{eqnarray}
where $\varepsilon_2=2 \varepsilon+2I$ is the energy of two $d$-fermions located at the nearest neighboring sites.

\begin{figure}[tp]
      \centering{\leavevmode}
      \begin{minipage}[h]{.49\linewidth}
\center{
\includegraphics[width=\linewidth]{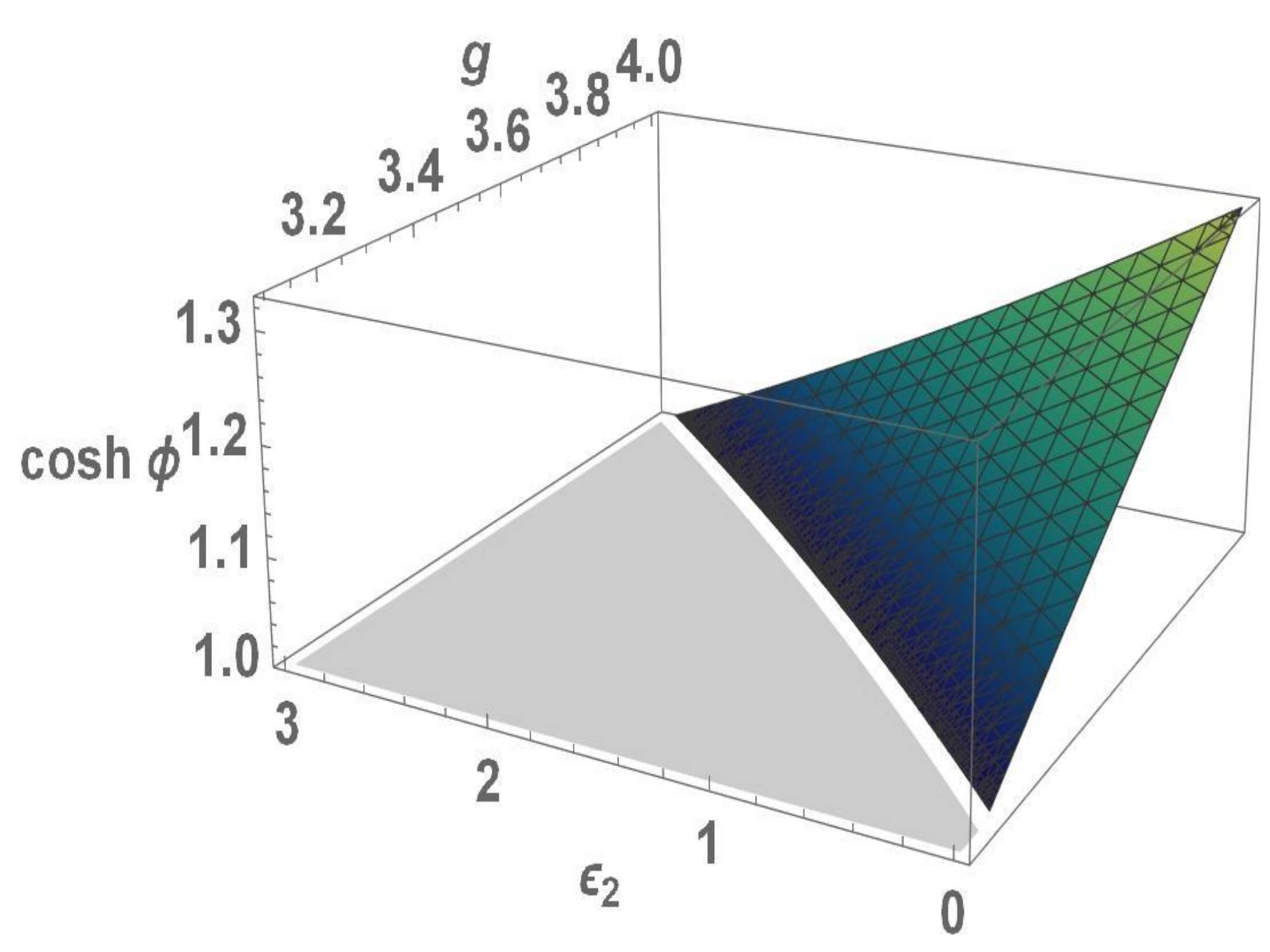} a)
                  }
    \end{minipage}
    \begin{minipage}[h]{.49\linewidth}
\center{
\includegraphics[width=\linewidth]{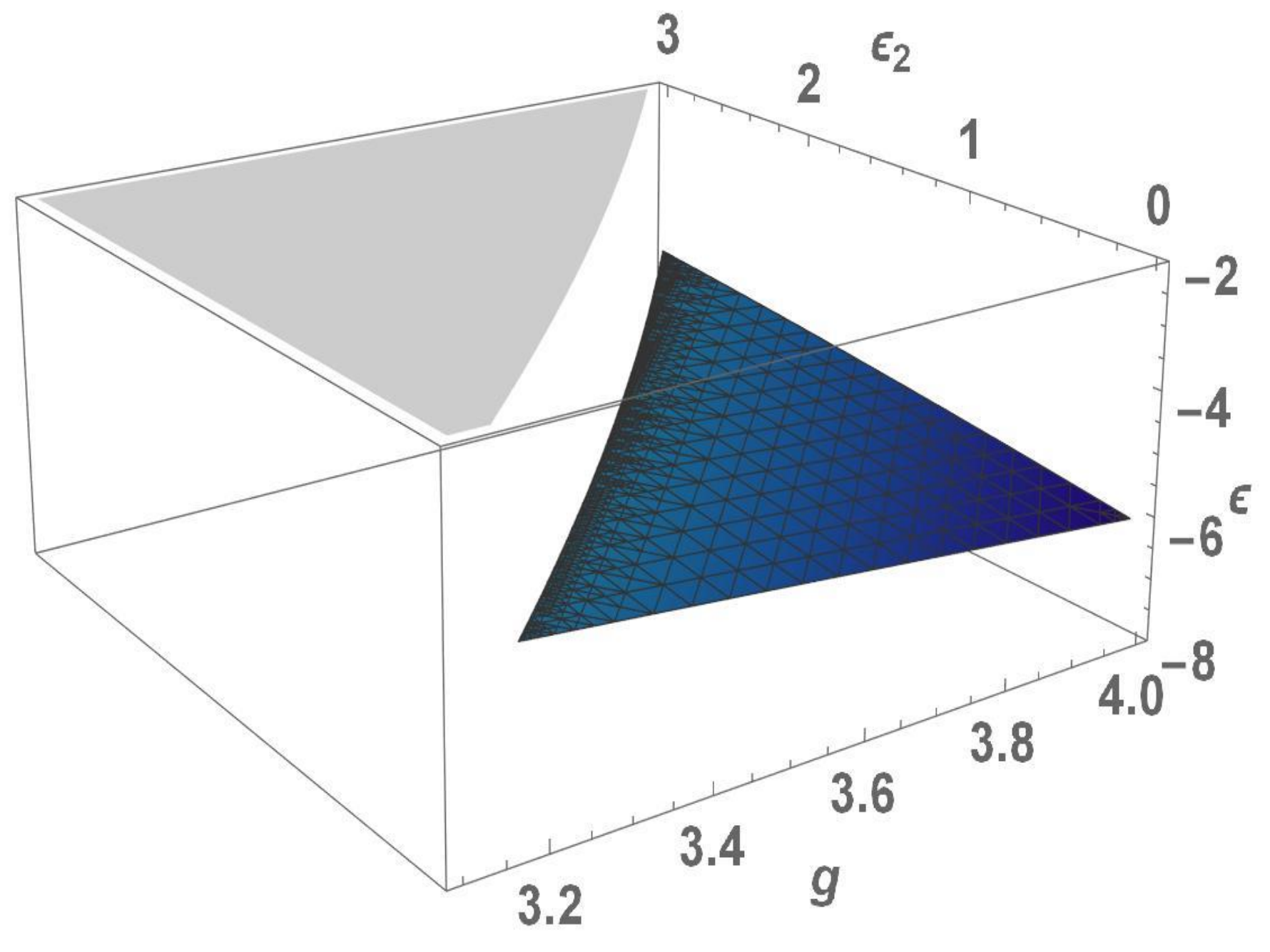} b)
                  }
    \end{minipage}
    \begin{minipage}[h]{.49\linewidth}
\center{
\includegraphics[width=\linewidth]{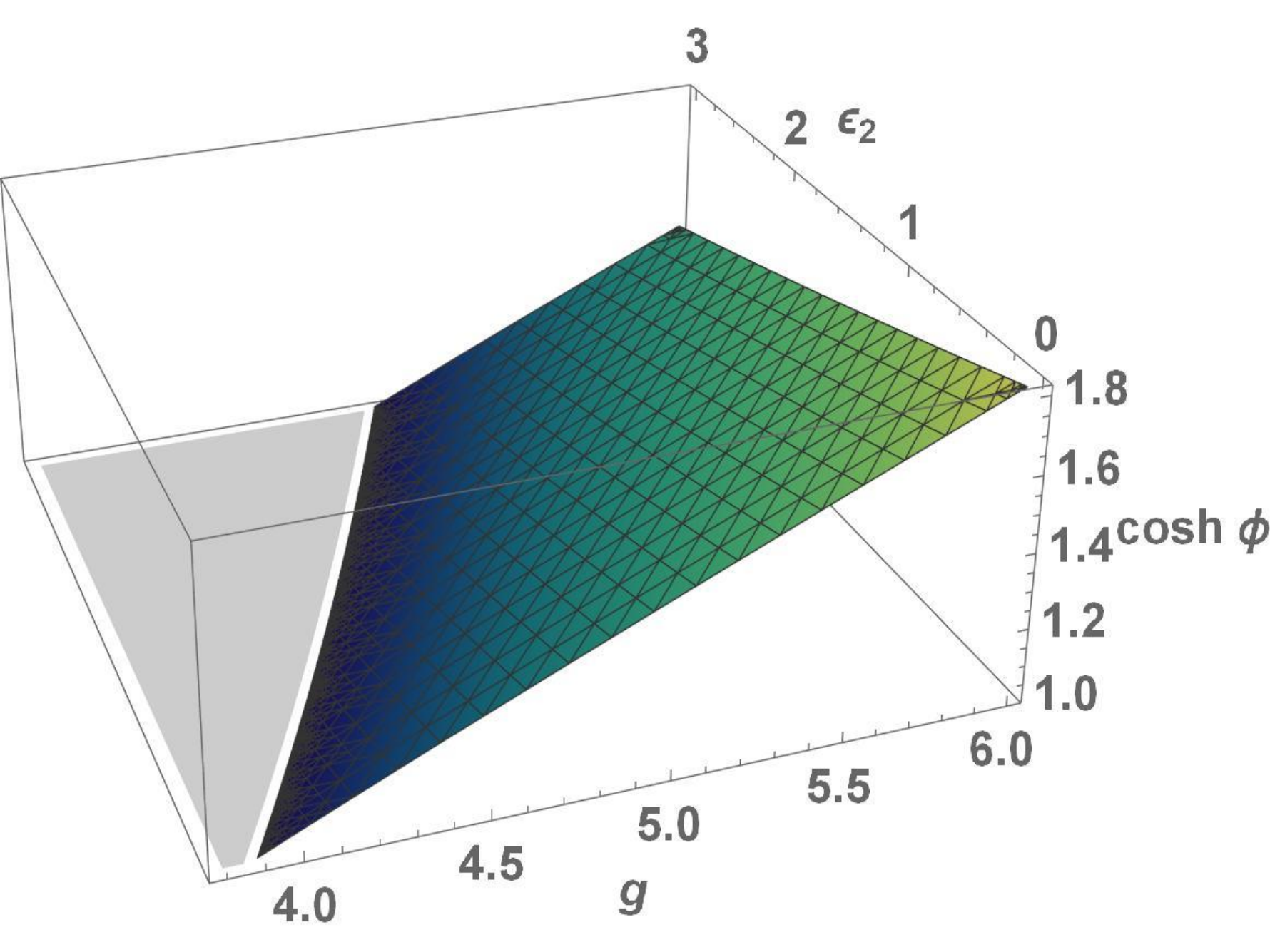} c)
                  }
    \end{minipage}
    \begin{minipage}[h]{.49\linewidth}
\center{
\includegraphics[width=\linewidth]{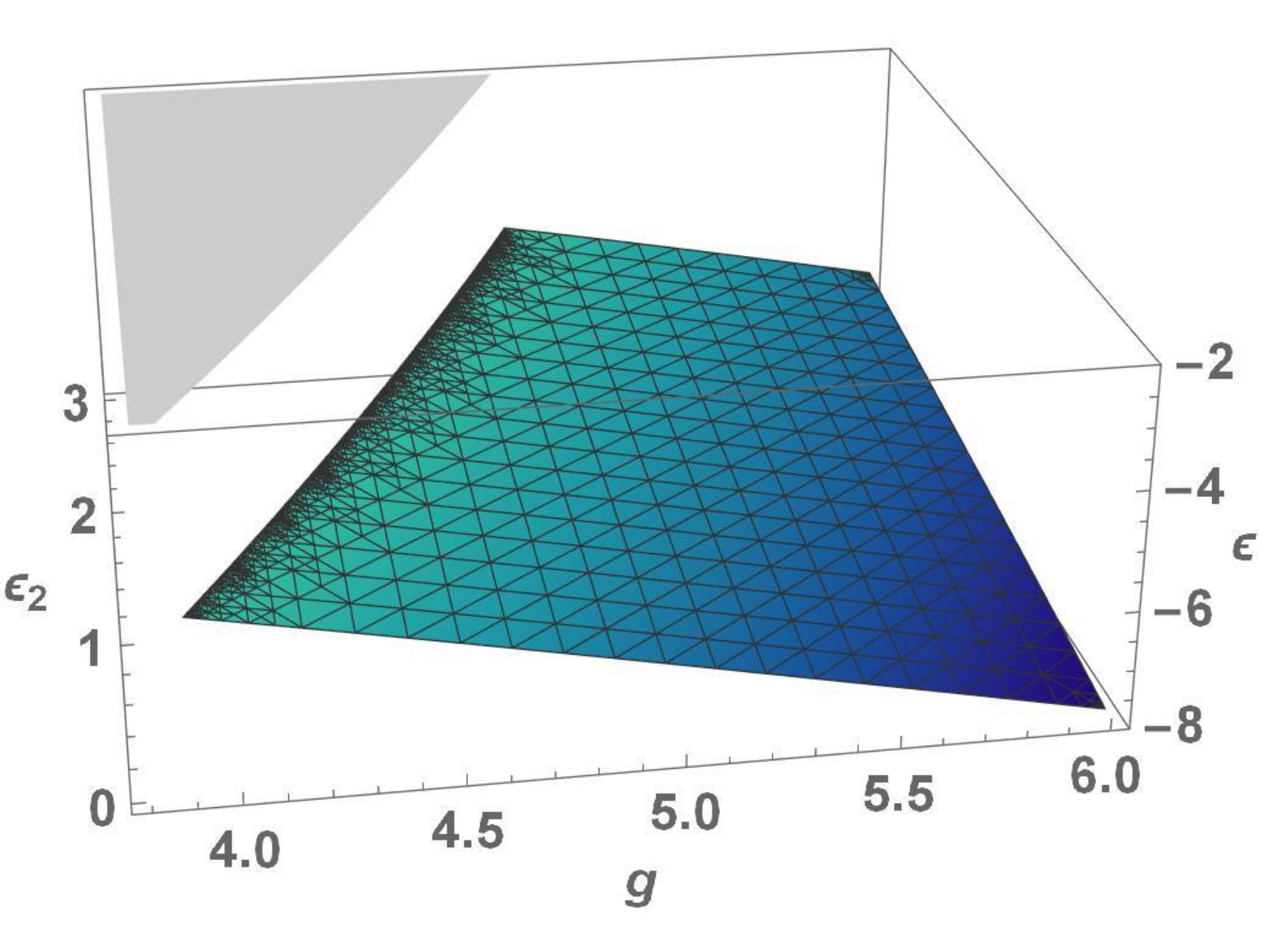} d)
                  }
 \end{minipage}
\caption{Effective attractive interaction between two $s$-fermions, forming a complex, $ J(k_1,k_2)=-\cosh \phi$ a),c) and energy of $s$-fermion pair $\epsilon$ b),d)
as function of $g$ and $\varepsilon_2$ calculated for $J=0$ a),b) and $J=\frac{1}{2}$ c),d).
} 
\label{fig:1}
\end{figure}

The solution for the two-particle wave function is determined by the Bethe ansatz in a traditional form 
$\psi (j_1,j_2)=A(12)\exp(ik_1j_1+ik_2j_2)+A(21)\exp(ik_1j_2+ik_2j_1),$ where the solution for the amplitudes $A(12)$ and $A(21)$,
determines a two-particle scattering matrix ${\cal S}(k_1,k_2)$ of $s$-fermions. 
The solution for the Bethe function follows from Eq.~(\ref{eq:5}), and the two-particle scattering matrix is defined as
\begin{eqnarray}
{\cal S}(k_1,k_2)=-\frac{\cos [(k_1+k_2)/2]+J(k_1,k_2)\exp[i (k_1-k_2)/2] }{\cos[(k_1+k_2)/2] + J(k_1,k_2)\exp[-i (k_1-k_2)/2])},
\label{eq:6}
\end{eqnarray}
where  $J(k_1,k_2)=J-\frac{g^2}{4\cos k_1+4\cos k_2 +2\varepsilon_2}$ defines an effective interaction between $s$-fermions.

To solve this problem, we will use periodic boundary conditions.  In the one-dimensional case, the Bette equations follow from the periodic boundary conditions for the Bethe function. The problem reduces to solving a set of $N$-coupled algebraic equations for the $N$
quasi-momenta $k_j$ of $s$-fermions, the Bethe equations for these unknown read
\begin{eqnarray}
&&\exp( ik_i L)=(-1)^{N-1}\prod_{j=1}^{N}S(k_i,k_j).
\label{eq:7}
\end{eqnarray}

We focus on studying the possibility of realizing bound fermion states \cite{IK1,IK0} based on solutions of the Bethe equations (\ref{eq:7}). 
In the case where an effective interaction between $s$-fermions $J_{eff}$ is a constant and  $J_{eff}<-1$, the complex solutions of rapidities $\lambda_j$ determine the ground state of the model. $\lambda_j$ is defined by the Orbach substitution \cite{O}: $\exp(ik_j)=\frac{\sin \frac{\phi}{2}(\lambda_j+ i)}{\sin\frac {\phi}{2}(\lambda_j- i)}$ with $\cosh \phi=-J_{eff}$  and $\phi >0$. In the thermodynamic limit solutions for rapidities are defined by complexes,  the complex is characterized by the abscissa $\alpha$ and its order $\nu$  $\lambda_\nu=\alpha+i\nu+i\delta_\nu$, $\nu=-(n-1),-(n-3),...(n-3),(n-1)$ ($\delta_\nu=\delta^*_\nu$, $\delta<<\frac{1}{L}$) \cite{G}. The  $\alpha$ value determines the total momentum, $\alpha=0$ corresponds to momentum  equal to zero. 

Consider a complex of two $s$-fermions, which is defined by the rapidities $\lambda=\pm 3i$.
The expressions for quasi-momenta $k_1=-k_2$ and $\exp ( i k_1)=2 \cosh \phi$ correspond to these solutions for rapidities. Taking into account the dependence of the effective interaction $J(k_1,k_2)$ on the quasi-momenta $k_1$ and $k_2$, we calculate the value of $\phi$ using the following equation
\begin{eqnarray}
&&\cosh \phi= -J+ \frac{g^2}{2} \frac{\cosh \phi}{4 \cosh^2 \phi +\cosh \phi \varepsilon_2 +1},
\label{eq:8}
\end{eqnarray}
at the condition $\phi >0$. The energy of  $s$-fermion pair is equal to  $\epsilon=-4\cosh \phi -\frac{1}{\cosh \phi}$.
 
Figs 2 a) and c) show calculations of the effective attractive interaction between $s$-fermions. Two-particle hybridization compensates for the repulsion between fermions in the case of strong interaction at $g>3$. The energy of $s$-fermion pair $\epsilon$ is shown in Figs 2 b),d). The energy of pairing fermion lies below the bottom of the conduction band $-2$ ($\epsilon<-4$).

Let us consider the simplest approximation, namely, we will take into account the states of $s$-fermions near the Fermi energy ($\varepsilon_F$) at $\varepsilon_2>0$. In this case, the magnitude of the effective interaction does not depend on the momenta of the fermions, is a constant and equal to $J(k_1,k_2)=J_{eff}=J-\frac{g^2}{2\varepsilon_2-4\varepsilon_F }$. In the case when the magnitude of the interaction does not depend on the momenta of fermions, 
 $|J_{eff}|=1$ is a special point for solutions for rapidities $\lambda_\nu$ of the Bethe equation (\ref{eq:7}) \cite{G}. The N-order complex corresponds to the Bose condensation of $s$-spinless fermions, since the usual pairing of two $s$-fermions is replaced by that for all fermions.
 As will be shown in the next section of the article, $\eta$-pairing of spinless fermions is realized under a soft condition. Thus, $\eta$-pairing is realized under the condition $g>g_\eta=\sqrt{2 J\varepsilon_2}$, and the Bose condensate is formed in the case  $g>g_{BC}=\sqrt{2(J+1)(\varepsilon_2-2\varepsilon_F)}$, thus $g_\eta <g_{BC}$, a value of $g_{BC}=\sqrt{2(J+1)\varepsilon_2}$ is reached at half-filling at $\varepsilon_F=0$.
 This is understandable since in this case we are actually talking about a condensate formed by complexes of two fermions for $\eta$-pairing.
 
\section{$\eta$-pairing}

In the model (\ref{eq:1}) $\eta$-pairing \cite{Y,6} is determined by two pairing operators $\eta=\sum_\textbf{j}\eta_\textbf{j}$, 
 $\eta_\textbf{j}=\exp (i \overrightarrow{\pi} \overrightarrow{j})c^\dagger_{\textbf{j}}c^\dagger_{\textbf{j+1}}$  and 
$\mu=\sum_\textbf{j} \mu_{\textbf{j}}$, $\mu_{\textbf{j}}=\exp (i \overrightarrow{\pi} \overrightarrow{j})d^\dagger_{\textbf{j}} d^\dagger_{\textbf{j+1}}$ for $s$- and $d$-fermions \cite{IK2,IK3}. The pairing operators are defined as sums over all lattice sites, for symmetric sample boundaries from $-\frac{L}{2}$ to $\frac{L}{2}$,  they must be symmetric with respect to $j$.
The coordinate-symmetric operators $\eta_\textbf{j}$ and $\mu_\textbf{j}$  determine non-trivial solutions $\eta$ and $\mu$.
These operators describe $s$- and $d$-fermion pairs, as these states are bounded  together through interaction. The pairing operators satisfy the following relations
\begin{eqnarray}
&&[{\cal H},\eta]=2J\eta-g\mu+2JD_{\eta}+g G_{\mu}-g \delta G_{\mu},\nonumber \\ 
&&[{\cal H},\mu]=\varepsilon_2 \mu  -g\eta +2I D_{\mu}+ g G_{\eta}-g \delta G_{\eta},
\label{eq:A}
\end{eqnarray}
where $D_{\eta}=\sum_{\textbf{j}}\eta_{\textbf{j}}(n^s_{\textbf{j+1}}+n^s_{\textbf{j+2}})$,
$D_{\mu}=\sum_{\textbf{j}}\mu_{\textbf{j}}(n^d_{\textbf{j+1}}+n^d_{\textbf{j+2}})$,
$G_{\mu}=\sum_{\textbf{j}}\mu_{\textbf{j}}(n^s_{\textbf{j}}+n^s_{\textbf{j+1}})$,
$G_{\eta}=\sum_{\textbf{j}}\eta_{\textbf{j}}(n^d_{\textbf{j}}+n^d_{\textbf{j+1}})$,
$\delta G_{\mu}=\sum_{\textbf{j}}\mu_{\textbf{j}}(c^\dagger_{\textbf{j+1}}c_{\textbf{j-1}}+c^\dagger_{\textbf{j}}c_{\textbf{j+2}})$ and $\delta G_{\eta}=\sum_{\textbf{j}}\eta_{\textbf{j}}(d^\dagger_{\textbf{j+1}}d_{\textbf{j-1}}+ d^\dagger_{\textbf{j}}d_{\textbf{j+2}})$.

In section Methods we will show, that the sums $D_{\mu}$, $D_{\eta}$, $ G_{\eta}$, $ G_{\mu}$, $\delta G_{\eta}$ and $\delta G_{\mu}$  are equal to zero.
Taking into account that the terms $D_{\mu}$, $D_{\eta}$, $\ G_\eta, G_\mu, \delta G_\eta, \delta G_\mu$ are equal to zero, the following solutions for energies of fermion pairs follow from Eqs (\ref{eq:A}) $\epsilon_\pm =\frac {1}{2}[\varepsilon_2+2J \pm \sqrt{4g^2+(\varepsilon_2-2J)^2}]$. 
For $\eta$-pairing, the fermion band widths
do not enter directly into the equation for the pairing operators; they are determined by the interaction constants.
The solutions for energy of fermion pairs is universal for the Hamiltonians in which the interaction is determined by the two-particle hybridization between fermions of different bands (compare with  \cite{IK2,IK3}).
 With an increase in  the value of $g$, the energy of the $s$-fermion pair decreases (from value $\epsilon_-=2J$ for $g=0$) and for  $g>\sqrt{2J \varepsilon_2}$  it becomes negative ($\epsilon_-<0$  for $\varepsilon_2>2J$).
In the case of $\epsilon_-<0$, $s$-spinless fermions form pairs (or $\eta$-pairing), creating a $p$-superconducting state, in contrast to \cite{IK2,IK3}, where electron pairs form  $s$-superconducting state during $\eta$-pairing. The superconducting state is realised at $\varepsilon_2>2J$ and $g>\sqrt{2J \varepsilon_2}$, when  $\epsilon_-<0$. 

The one-pair eigenstate of the Hamilton is $\vert \Psi_1>=G \vert vac>$, $G=\frac{1}{\sqrt{1+\phi^2}}(\phi \eta + \mu)$ ${\epsilon_-}$ (here $\phi=\frac{\varepsilon_2- \epsilon_-}{g}$), the energy $M{\epsilon_-}$ corresponds to $M$-pairs state $\vert \Psi_M>=G^M \vert vac>$. Is this state the interaction is determined by the effective Hamiltonian
\begin{eqnarray}
{\cal H}_{eff}=\epsilon_{-}\eta^\dagger(\textbf{q}) \eta (\textbf{q}),
\label{eq:10}
\end{eqnarray}

Unlike a Cooper pair, the kinetic energy of a pair in $\eta$-pairing is zero.  When the fermion filling exceeds half ($\varepsilon_F>0$ and the Fermi momentum $k_F>\frac{\pi}{2}$),  upon  the addition of a fermion with the energy $\varepsilon (k)$ with $k>k_F$, the energy of the fermion pair, equal to $\varepsilon(k)+\varepsilon (\pi -k)+\epsilon_- =\epsilon_-$, decreases as this pair is formed. This instability leads to the formation of a fermion condensate with a high pair density. The energy gap in the fermion spectrum is determined by the value of $\epsilon_-$. 
In the case $\epsilon_- <0$ the critical temperature is calculated as $-\epsilon_-=4 T_c$ \cite{6}.
It cannot be ruled out that in cuprates, where the superconducting state is realised at the doping of holes per Cu atoms (p)  with the effective itinerant carrier density $n_{eff}=1+p$ (with a maximal value of the superconducting transition temperature around $p\sim 0.16$) \cite{p}, $\eta$-pairing occurs. $\eta$-pairing leads to a spatial distribution of charge, which is also characteristic of the electronic states of cuprates near the superconducting phase.

The off-diagonal long-range order of this quantum state is equal to
\begin{eqnarray}
&&<\Psi_M\vert \eta_{j_1} \eta^\dagger_{j_2}\vert \Psi_M>\sim \frac{\phi^2}{1+\phi^2}\frac{M}{2L}\left(1-\frac{M}{2L}\right),
\label{eq:11}
\end{eqnarray}
where  $\vert j_1-j_2\vert \to \infty$.

The conditions for the realisation of $\eta$-pairing in the spinless fermion model (\ref{eq:1}) (which corresponds to $p$-superconducting state) and  the model with two-particle on-site hybridization of electrons  \cite{IK2,IK3} (which corresponds to $s$-superconducting state) are identical, namely when the effective interaction between fermions is attractive. In particular, if $J=0$ (or $U_c=0$ in \cite{IK2,IK3}), $\eta$-pairing is realised for any value of the interaction constant $g$ for $\varepsilon_2>0$. According to the exact solutions of the 1D Hubbard model and the chain of spinless fermions with density-density interaction, complex solutions of rapidities for the Bethe equation, which determine the bound pairs, are realised under different conditions: in the Hubbard chain, when the on-site interaction is negative, in the spinless fermion chain, when the interaction is negative and its magnitude exceeds a quarter of the band width ($J_{eff}<-1$).
As noted above, the conditions for the realization of $\eta$-pairing are independent of the fermion bandwidth when an effective attractive interaction is formed.

The solutions of the Bethe equations, which describe the many-body complex in the case of an attractive interaction, define a many-body state corresponding to $\eta$-pairing and correspond to the ground state of the system.   In such a complex, fermions are frozen, in exactly the same way as pairs in the case of $\eta$-pairing.
It must be acknowledged that the $\eta$-pairing proposed by Yang \cite{Y} reduces to a two-particle problem, so equations (9) takes on a simple form. The criterion for the realisation of the $\eta$-pairing in this case boils down to the condition $\epsilon_-<0$ or $g>\sqrt{2J \varepsilon_2}$.
The results obtained suggest that $\eta$-pairing corresponds to the ground state calculated using the two-particle approach.

In the case of the strong two-particle hybridization between $s$- and $d$-fermions, the condition $g>2J$ can be achieved; therefore, $\eta$-pairing of spinless fermions could be realized under realistic conditions.

\section{Conclusions} 

An exactly solvable 1D model of spinless fermions is proposed. It takes into account $d$-localized and $s$-itinerant spinless fermions interacting via the two-particle hybridization between fermions located at the nearest neighboring sites. It is shown that two-particle hybridization reduces the repulsive effective interaction between $s$-fermions located at the nearest neighboring sites, and in the case of strong interaction, this interaction becomes attractive. A maximum value of the attractive interaction between itinerant spinless fermions is realized when they form an $\eta$-pairing state. 
In the case of strong interaction, $\eta$-pairing of $s$-fermions located at the nearest neighboring sites is realized in the 3D model, determining the $p$-type superconducting state. Given that the $\eta$-pairing mechanism occurs when the filling of itinerant electrons is more than half-filled, it could be assumed that this pairing mechanism takes place in cuprates, where the superconducting state arises upon hole doping. 
Our results provide new insights into the nature of the pairing mechanism that could be realized in high-temperature superconductors.

\section{Methods}

Let us redefine one-particle operators $c^\dagger_\textbf{j}=\sum_{\textbf{k}}c^\dagger(\textbf{k})\exp (i\textbf{kj})$, $d^\dagger_\textbf{j}=\sum_{\textbf{k}}d^\dagger(\textbf{k})\exp (i\textbf{kj})$ and two-particle operators $\mu_\textbf{j}=\sum_\textbf{k}\mu(\textbf{k}) \exp (i\textbf{kj}+i\overrightarrow{\pi} \overrightarrow{j})$, 
$\eta_\textbf{j}=\sum_\textbf{k}\eta(\textbf{k}) \exp (i\textbf{kj}+i\overrightarrow{\pi} \overrightarrow{j})$. We also use  the following relation for pairing operators
$\eta_{\textbf{j+1}}=-\sum_\textbf{k}\eta(\textbf{k}) \exp (i\textbf{kj}+i\overrightarrow{\pi} \overrightarrow{j})\exp(i\textbf{k1})$.
Such $\sum_{\textbf{j}}\mu_{\textbf{j}}c^\dagger_{\textbf{j-1}}c_{\textbf{j+1}} +\sum_{\textbf{j}}\mu_{\textbf{j+1}}c^\dagger_{\textbf{j}}c_{\textbf{j+2}} = 
 \sum_{\textbf{k}_1,\textbf{k}_2} \mu (\textbf{k}_2-\textbf{k}_1-\overrightarrow{\pi}) c^\dagger ({\textbf{k}_1)}c(\textbf{k}_2) \exp[-i(\textbf{k}_1+\textbf{k}_2)\textbf{1}]-\exp[i(\textbf{k}_2 -\textbf{k}_1 -2\textbf{k}_2)\textbf{1}]=0$ (in second sum we shifted $j\to j+1$), we  obtain that $\delta G_{\mu}=\delta G_{\eta}=0$.
 
 Consider the general case where the operator  $Q_j=A_{j+\gamma} B_{j+\delta}\eta_{j}$ is defined by arbitrary operators  $A_j$, $B_j$ and $Q_j=Q_{-j}$. The equations for pairing operators are defined by the sums over the lattice sites of such operators. Let us calculate these sums over $j$ for $Q_j$ and for $Q_{j+1}$. Comparing these expressions, we obtain $\sum_j Q_j=-\sum_j Q_{j+1}$ or $\sum_j Q_{j-1/2}=-\sum_j Q_{j+1/2}$, where
$\sum_j Q_j=\sum_{k_1,k_2} A(k_1)B(-k_2)\eta (k_2-k_1-\pi)\exp (ik_1\gamma - ik_2\delta )$ and   
$\sum_j Q_{j+1}=-\sum_{k_1,k_2} A(k_1)B(-k_2)\eta (k_2-k_1-\pi)\exp [i k_1(\gamma +1) -i k_2(\delta +1)+i k_2- ik_1)]$. 
Using  the translation operator on $\frac{1}{2}$-lattice constant $T$ we redefine this expression $\sum_{j}(T+T^{-1})Q_j=0$. In other words, the sum over $j$ of the symmetric $Q_j$-operator is zero.

\subsection*{Author contributions statement}
I.K. is the author of the manuscript
\subsection*{Additional information}
The author declares no competing financial interests. 
\subsection*{Availability of Data and Materials}
All data generated or analysed during this study are included in this published article.

\end{document}